\documentclass[conference]{IEEEtran}
\IEEEoverridecommandlockouts

\usepackage{hyperref}
\usepackage{amsmath}
\usepackage[english]{babel}
\usepackage{times}
\usepackage{epsfig}
\usepackage{relsize}
\usepackage{enumerate}	
\usepackage{epstopdf}	
\usepackage{array}		
\usepackage{import}		
\usepackage{bytefield}
\usepackage{fixme}
\usepackage{makecell}
\usepackage{tabulary}
\usepackage{xcolor}
\usepackage{graphicx}
\usepackage{algorithmic}
\usepackage{bm}
\usepackage{float}
\usepackage{hhline}
\usepackage{pdfpages}
\usepackage{siunitx}
\usepackage{enumitem}
\usepackage{caption}
\usepackage{subcaption}
\usepackage{xspace}
\usepackage{scalefnt}
\usepackage{booktabs}
\usepackage{tabularx}

\usepackage{tikz}
\usepackage[demo]{tikzpeople}
\usepackage{pgfplots}
\DeclareUnicodeCharacter{2212}{−}
\usepgfplotslibrary{groupplots,dateplot}
\usetikzlibrary{patterns,shapes.arrows}
\pgfplotsset{compat=newest}

\usetikzlibrary{shapes}
\usetikzlibrary{intersections}
\usetikzlibrary{calc}
\usetikzlibrary{optics}
\usetikzlibrary{arrows}
\usetikzlibrary{positioning}
\usetikzlibrary{decorations}
\usetikzlibrary{angles}
\usetikzlibrary{quotes}
\usetikzlibrary{arrows.meta,shapes.geometric,decorations.pathreplacing}

\addto\extrasenglish{

}

\newcolumntype{C}{>{\centering\arraybackslash}X}

\newcommand{\covid}{COVID-19\xspace}
\newcommand{\cwbl}{clapping, whistling, booing, and laughter\xspace}

\def\BibTeX{{\rm B\kern-.05em{\sc i\kern-.025em b}\kern-.08em
    T\kern-.1667em\lower.7ex\hbox{E}\kern-.125emX}}

\begin{document}

\title{Enabling Acoustic Audience Feedback in Large Virtual Events}

\author{
\IEEEauthorblockN{Tamay Aykut\IEEEauthorrefmark{1}, Markus Hofbauer\IEEEauthorrefmark{2}, Christopher Kuhn\IEEEauthorrefmark{2}, Eckehard Steinbach\IEEEauthorrefmark{2}, Bernd Girod\IEEEauthorrefmark{3}}
\IEEEauthorblockA{\IEEEauthorrefmark{1}Sureel, Palo Alto, California, USA, tamay@sureel.ai \\
                  \IEEEauthorrefmark{2}Technical University of Munich, Germany, \{markus.hofbauer, christopher.kuhn, eckehard.steinbach\}@tum.de \\
                  \IEEEauthorrefmark{3}Stanford University, Stanford, California, USA, bgirod@stanford.edu
                }
}

\maketitle

\begin{abstract}
    The \covid pandemic shifted many events in our daily lives into the virtual domain.
    While virtual conference systems provide an alternative to physical meetings, larger events require a muted audience to avoid an accumulation of background noise and distorted audio.
    However, performing artists strongly rely on the feedback of their audience.
    We propose a concept for a virtual audience framework which supports all participants with the ambience of a real audience.
    Audience feedback is collected locally, allowing users to express enthusiasm or discontent by selecting means such as clapping, whistling, booing, and laughter.
    This feedback is sent as abstract information to a virtual audience server.
    We broadcast the combined virtual audience feedback information to all participants, which can be synthesized as a single acoustic feedback by the client.
    The synthesis can be done by turning the collective audience feedback into a prompt that is fed to state-of-the-art models such as AudioGen.
    This way, each user hears a single acoustic feedback sound of the entire virtual event, without requiring to unmute or risk hearing distorted, unsynchronized feedback.
\end{abstract}

\begin{IEEEkeywords}
Audio Synthesis, Multimedia Conference System, Virtual Audience
\end{IEEEkeywords}

\section{Introduction}
\label{sec:introduction}

Since \covid first hit, many live performances moved to the virtual domain in addition to in-person events.
While initially a necessary subpar substitute, virtual events are now a unique new opportunity for performing artists such as comedians, actors, and musicians.
However, performing such events without instantaneous feedback is challenging and not comparable with on-site shows.
The audience in larger online meeting needs to be muted to avoid disturbing background noise.
Interactive audience feedback such as applause or laughter is therefore not possible.

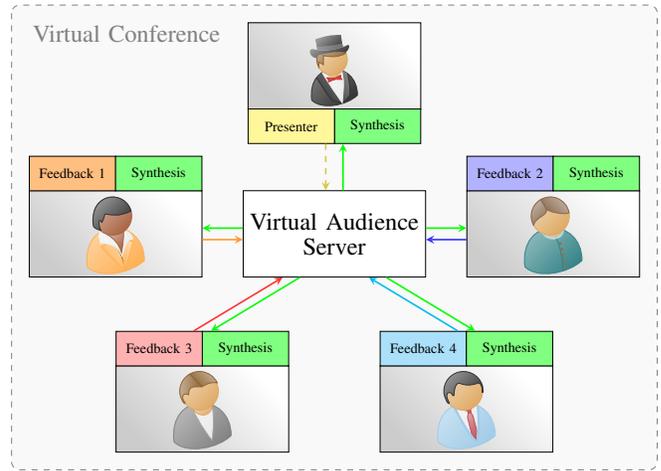
\begin{figure}[t]
    \centering
    \resizebox {\columnwidth} {!} {\begin{tikzpicture}

    \pgfdeclarelayer{background}
    \pgfsetlayers{background,main}

    \tikzset{node distance=4em and 2 em, >=stealth}
    \tikzset{root/.style= {align=center, minimum width = 3cm}}
    \tikzset{block/.style= {root, draw, rectangle, minimum height=1.5cm}}
    \tikzset{client/.style= {block, shading = axis, left color=gray!40, right color=white, shading angle=135}}
    \tikzset{synthesis/.style= {draw, below, fill=green!50, minimum width=1.5cm, minimum height=0.6cm, node distance= -0.015cm and -1.515cm}}
    \tikzset{interface/.style= {synthesis, above, fill=orange!50}}
    \tikzset{caption/.style= {synthesis, fill=yellow!50}}

    \node [block, fill=white] (server) {\large Virtual Audience\\ \large Server};
    \node [client, above = of server] (pres) {};
    \node [client, left = of server] (p1) {};
    \node [client, right = of server] (p2) {};
    \node [root, below = of server] (helper) {};
    \path (p1.east) |- (helper) node[client, midway, below] (p3) {};
    \path (p2.west) |- (helper) node[client, midway, below] (p4) {};

    \node[groom, minimum size=0.8cm, above = -1.45cm of pres] {};
    \node[alice, minimum size=1cm, above = -1.45cm of p1] {};
    \node[charlie, minimum size=1cm, above = -1.45cm of p2] {};
    \node[bob, minimum size=1cm, above = -1.45cm of p3] {};
    \node[dave, minimum size=1cm, above = -1.45cm of p4] {};

    \node[synthesis, below right = of pres] (pres_synth) {\scriptsize Synthesis};
    \node[caption, below left =  of pres](pres_caption) {\scriptsize Presenter};

    \node[interface, above left = of p1](p1_caption) {\scriptsize Feedback 1};
    \node[synthesis, above right = of p1](p1_synth) {\scriptsize Synthesis};

    \node[interface, above left = of p2, fill=blue!30](p2_caption) {\scriptsize Feedback 2};
    \node[synthesis, above right = of p2](p2_synth) {\scriptsize Synthesis};

    \node[interface, above left = of p3, fill=red!30](p3_caption) {\scriptsize Feedback 3};
    \node[synthesis, above right = of p3](p3_synth) {\scriptsize Synthesis};

    \node[interface, above left = of p4, fill=cyan!30](p4_caption) {\scriptsize Feedback 4};
    \node[synthesis, above right = of p4](p4_synth) {\scriptsize Synthesis};

    \path [draw, green, ->,thick] ($(server.west)+(0,0.1)$) -- ($(p1.east)+(0,0.1)$);
    \path [draw, orange!80,->,thick] ($(p1.east)+(0,-0.1)$) -- ($(server.west)+(0,-0.1)$);

    \path [draw, green, ->,thick] ($(server.east)+(0,0.1)$) -- ($(p2.west)+(0,0.1)$);
    \path [draw,blue!80, ->,thick] ($(p2.west)+(0,-0.1)$) -- ($(server.east)+(0,-0.1)$);

    \path [draw, green, ->,thick] ($(server.south)+(0.9,0)$) -- ($(p4_caption.north)+(0.9,0)$);
    \path [draw,cyan!80, ->,thick] ($(p4_caption.north)+(0.6,0)$) -- ($(server.south)+(0.6,0)$);

    \path [draw, green, ->,thick] ($(server.south)+(-0.6,0)$) -- ($(p3_synth.north)+(-0.6,0)$);
    \path [draw,red!80, ->,thick] ($(p3_synth.north)+(-0.9,0)$) -- ($(server.south)+(-0.9,0)$);

    \path [draw, green, ->,thick] ($(server.north)+(0.15,0)$) -- ($(pres_synth.south)+(-0.6,0)$);
    \path [draw, black!20!yellow,dashed, ->,thick] ($(pres_synth.south)+(-0.9,0)$) -- ($(server.north)+(-0.15,0)$);

    \begin{pgfonlayer}{background}
        \path (p1.west |- pres.north)+(-0.3,0.3) node (stl) {};
        \path (p2.east |- p4.south)+(0.3,-0.3) node (sbr) {};
        \path[fill=gray!5,rounded corners, draw=black!50, dashed] (stl) rectangle (sbr);
        \path (stl)+(2,-0.5) node [black!50]{\large Virtual Conference};
    \end{pgfonlayer}

\end{tikzpicture}}
    \caption{
        The proposed concept for a conference with virtual audience sound.
        A virtual audience server merges individual participation feedback, which is synthesized and played for each user.
    }
    \label{fig:motivation}
\end{figure}

So far, virtual conference systems do not offer the functionality to share acoustic audience feedback across the session.
Feedback by multiple participants at once is currently restricted to using text chat or signs such as hand waving.
This lack of audience interaction poses a significant challenge for performing artists who report that Zoom shows cannot replace in-person performances~\cite{vulture2020}.
It has been shown  that integrating laughter into virtual meetings significantly improves the social experience of the participants~\cite{niewiadomski_laugh-aware_2013}.
Without collective feedback of the audience such as laughter, a core part of human interaction in large events is missing.

Integrating acoustic audience feedback into an online conference faces several challenges.
Due to network lag, different feedback sounds need to be synchronized and normalized before playing them.
Raw audio data from many participants containing feedback such as laughter contains accumulating background noise.
An alternative to transmitting raw audio data is to synthesize it.
While the field of audio synthesis is mostly focused on speech synthesis~\cite{oord_wavenet_2016,prenger_waveglow_2018,engel_gansynth_2019}, nonverbal audio sounds such as laughter can be synthesized as well~\cite{tits_laughter_2020,urbain_avlaughtercycle_nodate}.
More recently, Meta's Audiocraft~\cite{copet2023simple} further increased the capabilities of audio generation based on abstract information in the form of text prompts.

In this paper, we propose to leverage recent advances in audio synthesis of acoustic audience feedback to integrate collective audience sound into virtual events.
In contrast to speech, information about the audience feedback such as laughter or clapping can be compressed efficiently into abstract state information.
We propose a virtual audience framework that allows audience interaction without the transmission of the actual audio information.
The concept of the proposed framework is shown in \autoref{fig:motivation}.

On the client input side, every participant shares abstract information about their reactions which we merge at a central virtual audience server.
We then use the abstract feedback state to synthesize a single joint audience sound, for example by turning the feedback state into a prompt and using text-to-sound generative models to obtain the audio.
The joint audio sound is then sent back to each user, allowing each user to hear the feedback of the entire audience without overlays or distortion.
By continuously updating the feedback state, the audio played for each user can be updated repeatedly to enable an audience sound that fits the current events.

The rest of this paper is organized as follows.
\autoref{sec:relatedWork} summarizes related work in the field of audience audio synthesis.
We propose a joint acoustic audience framework for large virtual events in \autoref{sec:framework}.
Then, we present a potential implementation of synthesizing joint audio sounds using state-of-the-art generative AI.
\autoref{sec:conclusion} concludes the paper.

\section{Related Work}
\label{sec:relatedWork}

In this section, we summarize techniques for synthesizing acoustic audience feedback.
Then, we give an overview of how integrating audience feedback in virtual events is handled currently.

\subsection{Sound Synthesis}

Traditional sound synthesis techniques can be separated into five categories: sample-based, physical modeling, signal modeling,abstract synthesis~\cite{smi91}, and learning-based synthesis.
More recently, deep-learning-based synthesis approaches have redefined the possibilities in sound synthesis~\cite{copet2023simple}. 

In \textit{sample-based synthesis}, audio recordings are cut and spliced together to produce new sounds.
The most common example of this is granular synthesis~\cite{roads1988introduction}.
A sound grain is generally a small element or component of a sound, typically between \SIrange{10}{200}{\milli\second} in length.
Concatenative synthesis is a subset of granular synthesis~\cite{simon2005audio}.
The goal is to select and recombine the grains in a way that avoids perceivable discontinuities.

Instead of using prerecorded audio data, \textit{physical modeling synthesis} aims to model the underlying physical process of a sound system.
Physical models require solving partial differential equations for each sample~\cite{bil09}.
The resulting models are computationally intensive, requiring significant GPU resources to run in real time~\cite{har18,web15}.

In \textit{signal modeling synthesis}, sounds are created based on an analysis of real-world sounds.
The analyzed waveform is then used for synthesis.
The most common method of signal modeling synthesis is Spectral Modeling Synthesis~\cite{ser90}.
Spectral modeling can be approached by analyzing the original audio file, selecting a series of sine waves to be used for synthesis, and then combining it with a residual noise shape to produce the original sound~\cite{am02}.

In \textit{abstract synthesis}, sounds are obtained using abstract methods and algorithms, typically to create entirely new sounds.
An example is Frequency Modulation (FM) synthesis~\cite{ch73}.
Two sine waves are multiplied together to create a more complex, richer sound that might not exist in the natural world.
Early video game sounds were often based on FM synthesis.
These sounds can be created and controlled in real-time due to the low complexity of the required process.

Finally, in \textit{deep-learning-based synthesis}, large amounts of recordings are used to obtain a sound synthesis model in a data-driven way~\cite{huzaifah_deep_2020,latif_deep_2020}.
Autoencoders have shown great promise for this task~\cite{roche_autoencoders_2019}, both for music~\cite{dhariwal_jukebox_2020} and speech synthesis~\cite{binkowski_high_2019}.
Architectures such as WaveNet~\cite{oord_wavenet_2016} allow to learn a generative synthesis model directly from real-world recordings, generating significantly more natural sounds than parametric systems.
While such models are complex and computationally expensive, recent architectures have increased the inference speed~\cite{ren_fastspeech_2020,yang_multi-band_2020,vainer_speedyspeech_2020}.
In 2023, Meta released Audiocraft, which includes text-to-sound systems such as AudioGen~\cite{copet2023simple} or MusicGen~\cite{kreuk2022audiogen}.
These models allow turning natural text into arbitrary sound, or into music.
For the proposed framework, this flexible language-based approach allows easily turning abstract audience feedback data into sound by turning the abstract data into a text prompt first.

\subsection{Acoustic Feedback Synthesis}

Next, we address specific implementations of sound synthesis for creating the most common acoustic feedback sounds of \cwbl.

Since the physical mechanism of \textit{Clapping} is straightforward, synthesizing clapping sounds can be approached using physical modeling~\cite{peltola2007synthesis}.
\textit{Whistling} can be approached using abstract FM synthesis~\cite{mori2010method}.
\textit{Booing} can be generated using both abstract or sample-based synthesis~\cite{cardle2003audio}.

The most complex and challenging sound to synthesize in a virtual audience is \textit{Laughter}.
Since an individual laughter is already a complex sound and additionally varies significantly from person to person, the most promising approaches for laughter synthesis are based on deep learning.
Mori~et~al.~\cite{mo19} used WaveNet~\cite{oord_wavenet_2016} to generate synthetic laughter that outperformed laughter synthesized using Hidden Markov Models (HMMs).
They used the Online Gaming Voice Chat Corpus~\cite{arimoto_naturalistic_2012} to condition WaveNet, allowing them to control the amplitude envelope of the synthesized laughter.
Despite the improved naturalness, the resulting laughter was still largely perceived as noisy and echoic.

Another approach used transfer learning from models trained with speech data to circumvent the problem of lack of laughter training data~\cite{tits_laughter_2020}.
First, text-to-speech is trained and then fine-tuned with smiled speech and laughter data.
MelGAN~\cite{kumar2019melgan} is used to obtain the output waveform.

Finally, generative artificial intelligence methods such as Generative Spoken Language Modeling~\cite{lakhotia2021generative} or AudioGen~\cite{copet2023simple} can be used to generate laughter from text prompt.
Fine-tuning these models specifically for laughter is a promising direction for natural-sounding laughter synthesis.

\section{Virtual Audience Framework}
\label{sec:framework}

In this section, we present the proposed virtual audience framework.
We focus on online events where a single client provides the content for all other participants in the session.
This single presenting client interacts with the audience based on their reactions and uses this as feedback for their future behavior.
Such feedback is especially relevant for performing artists that rely on live shows.

\subsection{Concept}

In the scenario we focus on, the presenter depends on the reaction of the entire audience rather than detailed feedback by individuals.
In case of multiple participants talking or the entire audience sending feedback at the same time, noise can accumulate and drown out valuable feedback.
To avoid accumulating noise, the audio of multiple participants in an online meeting typically needs to be synchronized.
Since nonverbal acoustic feedback information such as \cwbl is less complex than speech, we propose to avoid requiring complex synchronization schemes.
We only require an abstract representation of the current audience state denoting which participant is actively \cwbl.
This abstract audience state can then be shared with all participants.
The acoustic feedback can then be synthesized locally from the received audience state.
This way, each participant can be played the overall acoustic feedback without having to synchronize raw audio.

\subsection{Implementation}

\begin{figure}[ht!]
    \centering
    \resizebox {\columnwidth} {!} {\begin{tikzpicture}

    \pgfdeclarelayer{background}
    \pgfsetlayers{background,main}

    \tikzset{node distance=2em and 2 em, >=stealth}
    \tikzset{root/.style= {align=center, text width = 3cm}}
    \tikzset{block/.style= {root, draw, rectangle, minimum height=1.5cm, fill=blue!30}}
    \tikzset{client/.style= {block, fill=orange!30}}
    \tikzset{synthesis/.style= {draw, below, fill=green!50}}
    \tikzset{interface/.style= {draw, above, fill=orange!50}}

    \node [block] (server) {Audience State Management};
    \node [block, left = of server] (input) {Audience State Feedback};
    \node [block, right = of server] (output) {Audience State Broadcasting};
    \node [client, below = of input] (pres) {Presenter};
    \node [client, right = of pres] (p1) {Participant 1};
    \node [client, right = of p1] (pn) {Participant N};

    \path [draw, ->] (input) -- (server);
    \path [draw, ->] (server) -- (output);

    \path [draw, ->] (output.south) -- (pn.north) node[synthesis] {\scriptsize Audio Synthesis};
    \path [draw, ->] (output.south) -- +(0,-0.4) -| (p1.north) node[synthesis] {\scriptsize Audio Synthesis};
    \path [draw, ->] (output.south) -- +(0,-0.4) -| (pres.north) node[synthesis] {\scriptsize Audio Synthesis};

    \node [left = of pres] (input-helper) {};
    \path [draw, ->] (pres.south) node[interface] {\scriptsize Audience State Interface} -- +(0,-0.4) -| (input-helper.center) |- (input.west);
    \path [draw, ->] (p1.south) node[interface] {\scriptsize Audience State Interface} -- +(0,-0.4) -| (input-helper.center) |- (input.west);
    \path [draw, ->] (pn.south) node[interface] {\scriptsize Audience State Interface} -- +(0,-0.4) -| (input-helper.center) |- (input.west);

    \begin{pgfonlayer}{background}
        \path (input.west |- input.north)+(-0.3,0.6) node (stl) {};
        \path (output.east |- output.south)+(0.3,-0.2) node (sbr) {};
        \path[fill=gray!20,rounded corners, draw=black!50, dashed] (stl) rectangle (sbr);
        \path (server)+(0,1.1) node {Virtual Audience Server};
    \end{pgfonlayer}

\end{tikzpicture}}
    \caption{
        Overview of the proposed virtual audience server which collects audience feedback from each user.
        The feedback is combined and sent back to every client.
        On the client, we synthesize the overall audience sound locally.
    }
    \label{fig:overview}
\end{figure}
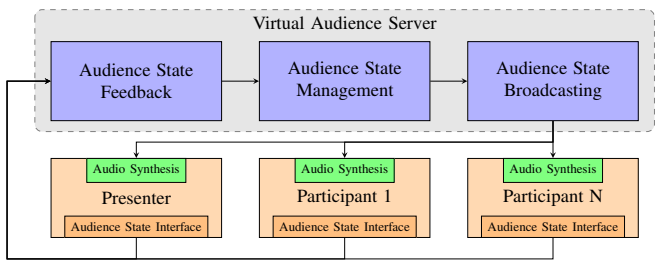

To implement the proposed concept, we share the abstract audience state information with all participants via a central server.
We define the audience state as a vector of binary variables for each participant, with each variable representing whether the user is exhibiting a particular reaction.
Whenever a client changes their reaction state, the updated state information is sent to the central virtual audience server.
The central virtual audience server merges all information and broadcasts the updated current audience state information to every client.
Finally, every client synthesises the audience feedback locally for the current audience state.
The transmission of such abstract state information results in an additional transmission rate of a few bytes which is fundamentally less compared to the transmission of an audio or video stream.
\autoref{fig:overview} summarizes the concept for the virtual audience framework.

On the client side, we define an input interface that collects the abstract state information from the users and an output interface that converts the current audience state information into an audio stream.
For synthesizing the audio, any state-of-the-art technique such as AudioGen~\cite{copet2023simple} or MusicGen~\cite{kreuk2022audiogen} can be used.

The input interface enables a flexible implementation due to the simplicity of the abstract state information that will be transmitted to the virtual audience server.
A Graphical User Interface (GUI) is a simple way for allowing audience contributions.
Similar to hand waving in Zoom, participants can share feedback via their GUI.
This feedback is collected on the sever and the resulting audience state will be synthesized on every client into an acoustic audience signal.
We obtain the abstract state information from the \cwbl of the user using a GUI or detection methods such as \cite{hagerer_did_nodate}.
The transmission of the abstract state information instead of the actual audio signal avoids audio synchronization issues and comes with a negligible transmission overhead as small as a few bytes.

\section{Conclusion}
\label{sec:conclusion}

In this paper, we presented a virtual audience framework for online conferences.
Performers such as actors, comedians, or musicians rely heavily on the feedback of their audience.
This work addressed the issue of accumulating noise caused by multiple audio inputs that so far is being solved by requiring the audience to be muted.
The proposed virtual audience framework enables all participants to experience the audience feedback without the transmission of an audio stream and the resulting synchronization issues.
We collect abstract audience state information, such as the number of clapping and laughing participants, on a central server and synthesize a unified audience sound locally on every client.
Every user contributes to the overall audience state and has direct influence on the synthesized audio information.

In future work, reactions such as laughter can be locally detected using methods such as deep neural networks~\cite{hagerer_did_nodate} to then generate the abstract audience state data which will be shared with the server.
Furthermore, the abstract state information is not restricted to be binary.
The field of audio synthesis offers promising ideas such as acoustic unit discovery~\cite{eloff_unsupervised_2019,chorowski_unsupervised_2019}.
The acoustic units present in the acoustic feedback of an audience member can be used as a more informative state data. The joint audience sound can consist of the same abstract units to achieve a sound that closely resembles the actual sound.
Such improved synthesis implementations can be easily added to the proposed modular framework to continually improve the virtual audience sound synthesis.

\section{Acknowledgement}

This work has been supported by the Max Planck Center for Visual Computing and Communication.

\bibliographystyle{IEEE}
\bibliography{IEEEabrv,references}

\end{document}